# PROBING FOR MACHOS OF MASS $10^{-15} M_\odot$–$10^{-7} M_\odot$ WITH GAMMA-RAY BURST PARALLAX SPACECRAFT

ROBERT J. NEMIROFF[1]

NASA Goddard Space Flight Center, Code 668.1 Greenbelt, MD 20771
Internet: nemiroff@grossc.gsfc.nasa.gov

ANDREW GOULD[2]

Dept. of Astronomy, Ohio State University, Columbus, OH 43210
Internet: gould@payne.mps.ohio-state.edu

## Abstract

Two spacecraft separated by $\sim 1\,\mathrm{AU}$ and equipped with gamma-ray burst (GRB) detectors could detect or rule out a cosmological density of Massive Compact Halo Objects (MACHOs) in the mass range $10^{-15} M_\odot < M < 10^{-7} M_\odot$ provided that GRBs prove to be cosmological. Previously devised methods for detecting MACHOs have spanned the mass range $10^{-16} M_\odot < M < 10^7 M_\odot$, but with a gap of several orders of magnitude near $10^{-9} M_\odot$. For MACHOs and sources both at a cosmological distance, the Einstein radius is $\sim 1\,\mathrm{AU}\,(M/10^{-7} M_\odot)^{1/2}$. Hence, if a GRB lies within the Einstein ring of a MACHO of mass $M < 10^{-7} M_\odot$ as seen by one detector, it will not lie in the Einstein ring as seen by a second detector $\sim 1\,\mathrm{AU}$ away. This implies that if GRBs are measured to have significantly different fluxes by the two detectors, this would signal the presence of a MACHO $<10^{-7} M_\odot$. By the same token, if the two detectors measured similar fluxes for several hundred events a cosmological abundance of such low-mass MACHOs would be ruled out. The lower limit of sensitivity, $M < 10^{-15} M_\odot$ is set by the finite size of the source. If low-mass MACHOs are detected, there are tests which can discriminate among events generated by MACHOs in the three mass ranges $M < 10^{-12} M_\odot$, $10^{-12} M_\odot < M < 10^{-7} M_\odot$, and $M > 10^{-7} M_\odot$. Further experiments would then be required to make more accurate mass measurements.

# 1. INTRODUCTION

While there is no *a priori* reason to expect that Massive Compact Halo Objects (MACHOs) in the mass range $10^{-15} M_\odot < M < 10^{-7} M_\odot$ comprise a significant fraction of the density of the universe, neither is there any definitive argument ruling them out. There are three main candidates for MACHOs in this mass range: snow balls, black holes, and small molecular clouds. Snow balls are lumps of primordial baryons (H and He). Snow balls in this mass range are generally believed to evaporate on time scales short compared to the Hubble time (De Rújula, Jetzer & Massó 1992) and for this reason are often dismissed. However, the argument for evaporation might not be considered definitive. Primordial black holes in this mass range would not have had time to evaporate by Hawking radiation, and generally would not radiate enough to be detected. On the other hand, there are no detailed scenarios in which significant numbers of primordial black holes in this mass range would form. Small molecular clouds have recently been advanced as a solution or partial solution to the dark matter problem, if for no other reason than they are extremely hard to detect (Pfenniger, Combes, & Martinet 1994). Small molecular clouds are not usually grouped as MACHOs since they are diffuse rather than "compact". However for MACHOs the word "compact" in "Massive Compact Halo Object" does not refer to any specific density. It simply means that the object is compact enough to fit into its own Einstein ring. For Galactic MACHOs, this size is $r_e < 1 \, \text{AU} \, (M/0.05 \, M_\odot)^{1/2}$, thus excluding molecular clouds. For cosmological MACHOs, much larger radii are permitted: $r_e < 1 \, \text{AU} \, (M/10^{-7} \, M_\odot)^{1/2}$ which would include many small molecular clouds. Small molecular clouds are expected to have a fractal structure implying that they could be probed on small scales by GRB parallax measurements. (In this *Letter* we use to word 'MACHO' to refer to any class of massive compact objects, whether or not they literally live in the halos of galaxies.)

The mass range probed by GRB parallaxes is quite a good scale for "hiding" dark matter, even if the candidate objects have not previously received a great deal of attention. Baryonic objects of this mass would be too cold to emit much light of their own - they would be truly "dark." In particular, no methods have previously been developed for probing the mass scale near $10^{-9} M_\odot$ because the objects are so dark and because this range had been thought to be inaccessible to gravitational lensing (Nemiroff 1993).

Here we show that GRB detectors aboard two spacecraft separated by $\sim 1 \, \text{AU}$ could detect or rule out low-mass MACHOs. Refsdal (1966) was the first to point out that by observing a MACHO event from two platforms separated by solar-system scale distances, one could obtain significantly more information than from a single platform. Recently this idea of a "parallax spacecraft" has been suggested as a method of better constraining the velocities and masses of Galactic microlensing events (Gould 1992b, 1994, 1995a; Drukier, Nemiroff, & Ozernoy 1994) and determining the transverse velocities of galaxies (Grieger,

Kayser, & Refsdal 1986; Gould 1995b).

Our proposal requires that GRBs occur at cosmological distances. For a discussion on the arguments for and against this hypothesis see Paczyński (1995) and Lamb (1995). Paczyński (1986, 1987) was the first to point out that if they are at such great distances GRBs could undergo a detectable microlensing effect. Mao & Paczyński (1992) estimated the chance that GRBs might undergo a galactic (macro-)lensing effect and Blaes & Webster (1992) suggested a that cosmological abundance of MACHOs $M \sim 10^6\,M_\odot$ could be found by searching for auto-correlations in the time series of measured GRB fluxes. Nemiroff et al. (1993) have excluded a closure density of MACHOs between $10^{6.5}$ and $10^{8.1} M_\odot$ for a conservative estimate of GRB redshifts, with dim bursts lying near a redshift of unity. Gould (1992a) suggested the possibility of observing femtolensing of GRBs by MACHOs $10^{-16} M_\odot < M < 10^{-13}\,M_\odot$ from the frequency dependent interference pattern in the GRB spectrum and Ulmer & Goodman (1995) discussed post-WKB femtolensing effects which may allow detection of MACHOs as heavy as $10^{-11}\,M_\odot$.

Our basic idea is extremely simple. For MACHOs and sources at a cosmological distance, and for MACHOs of sufficiently low mass ($M < 10^{-7}\,M_\odot$), the size of the Einstein ring projected onto the solar system will be $\tilde{r}_e < 1\,\mathrm{AU}$. In this case were the source to lie within the Einstein ring as seen by one spacecraft, it would not lie within the Einstein ring as seen by a second spacecraft $\sim 1\,\mathrm{AU}$ away. The signature of a lensing event is simply that a single GRB is observed to have significantly different fluxes as seen from the two spacecraft.

In § 2 we discuss this idea quantitatively and show that it is sensitive to a range of mass $10^{-15}\,M_\odot < M < 10^{-7}\,M_\odot$. If no events were detected, this would place limits on the cosmological density of such objects. If some events were detected, one would know only the total optical depth of MACHOs within the mass range to which the experiment is sensitive, but one would not generally have any additional information constraining the mass. In § 3, we discuss how such additional information might be obtained. For MACHOs near the lower end of the range, the ratio of the fluxes at the two spacecraft will vary with time, permitting a rough estimate of the mass. For MACHOs near the upper end, the distribution of flux ratios measured in different events will give a clue to the mass. We also discuss possible follow-up experiments which could provide additional information about $M$.

## 2. ANALYSIS

A lens magnifies a point source by different amounts as seen by different observers. An observer directly in line with the lens and the source would detect a ring of formally infinite magnification: the Einstein ring (Einstein 1936). The observer sees the Einstein ring as having some angular radius $\theta_e$, and, given the distance to the lens, can compute the radius $r_e$ of the Einstein ring in the lens plane. An observer for whom the source and lens were separated by an angle $\theta$ would see two images of the source with combined magnification,

$$A(x) = \frac{x^2 + 2}{x(x^2 + 2)^{1/2}} \qquad x \equiv \frac{\theta}{\theta_e}. \tag{1}$$

The physical Einstein radius $r_e$ can be projected onto either the observer or the source plane yielding two new length scales $\tilde{r}_e$ and $\hat{r}_e$. If two observers are separated by $\tilde{r}_e$, then the unlensed source positions appear displaced by $\theta_e$. Two sources separated by $\hat{r}_e$ have unlensed angular separations $\theta_e$. The four quantities so defined are related by

$$\theta_e = \sqrt{\frac{2\,R_S\,D_{LS}}{D_{OL}D_{OS}}} = \frac{r_e}{D_{OL}} = \frac{\hat{r}_e}{D_{OS}} = \frac{\tilde{r}_e D_{LS}}{D_{OL}D_{OS}}, \tag{2}$$

where $D_{OL}$, $D_{OS}$, and $D_{LS}$, are the distances between the observer, lens, and source, and $R_S$ is the Schwarzschild radius of the lens: $R_S = 2GM/c^2 = 3\,\text{km}(M/M_\odot)$. In a cosmological setting, the distances discussed are all angular diameter distances (see e.g. Turner, Ostriker, & Gott 1984). The generic term "Einstein ring" when used in the literature usually refers to $r_e$. Note that this quantity is related to $\tilde{r}_e$ as absolute to relative parallax.

Let $v$ be the transverse speed of the lens relative to the observer-source line of sight and let $t_b$ be the duration of the GRB. Then the magnification will be essentially constant during the GRB provided

$$\frac{vt_b}{r_e} \ll 1. \tag{3}$$

We initially assume that equation (3) holds. In addition, we assume that the event is detected by two observers separated by a distance $a$ that is large compared to the projected Einstein ring, $\tilde{r}_e/a \ll 1$. It follows immediately that if the GRB is within the Einstein ring as seen by first observer ($x_1 < 1$), then it will be very far from Einstein ring as seen by the other ($x_2 \gg 1$). Hence, the ratio of the fluxes as seen by the two observers (which is equal to the ratio of magnifications $\mathcal{R} \equiv A_1/A_2$) is just equal to $A_1$. For a given event the probability that an observer lies inside the Einstein ring of some MACHO is by definition $\tau$. Let $f(\mathcal{R})d\mathcal{R}$ be the probability that the observed ratio of fluxes lies between $\mathcal{R}$ and $\mathcal{R} + d\mathcal{R}$. And let $F(\mathcal{R}) \equiv \int_\mathcal{R}^\infty d\mathcal{R}' f(\mathcal{R}')$. From equation (1) one then finds that

$$f(\mathcal{R}) = 2\tau[\mathcal{R}^2 - 1]^{-3/2}, \qquad F(\mathcal{R}) = 2\tau[(1 - \mathcal{R}^{-2})^{-1/2} - 1]. \tag{4}$$

The ratio $\mathcal{R}^\dagger \equiv A_2/A_1$ has the same distribution.

Consider then an experiment which is sensitive to flux ratios which differ from unity by at least $\mathcal{R}_{\min}$. For definiteness, we take $\mathcal{R}_{\min} = 1.34$ corresponding to $x_1 < 1$. And assume that $N$ GRBs are observed. Then one would expect $2N\,F(\mathcal{R}_{\min}) = 2N\tau$ events where one or the other observer saw significantly more flux than the other. Hence, if no such events were observed, one could rule out optical depths $\tau > 1.5/N$ at the 95% confidence level.

The expected optical depth is given by

$$\tau = g\Omega_L, \qquad (5)$$

where $\Omega_L$ is the density of lenses in units of the closure density of the universe. While the parameter $g$ depends on distribution of the sources and to a lesser extent on the geometry of the universe, one finds that for a mean source redshift of $z \sim 0.5$ appropriate under one set of estimates for the brighter GRBs (Wickramasinghe et al. 1993; Norris et al. 1994; Cohen, Kolatt, and Piran 1995) $g \sim 0.05$ (Gould 1992a; Nemiroff et al. 1993). This means that with only a few hundred events one could probe to densities $\Omega_L \sim 0.1$, the value characteristic of dark matter on galactic scales.

Let us assume for the moment that indeed no events were detected with flux ratios significantly different from unity. What range of masses would be ruled out? The upper mass limit is set by the assumption that the two observers are separated by more than an Einstein ring. The most probable scenario for a detectable lensing event occurs when the lens distance is a reasonable fraction of the source distance (Turner, Ostriker, and Gott 1984; Nemiroff 1989). From equation (2), $r < \sqrt{R_S D_{OS}}$. For definiteness, we adopt $D_{OS} \sim 1\,\mathrm{Gpc}$ and find

$$M < 10^{-7} M_\odot \left(\frac{r}{\mathrm{AU}}\right)^2 \left(\frac{D_{OS}}{\mathrm{Gpc}}\right)^{-1}. \qquad (6)$$

The lower mass limit is set by the assumption codified in equation (3) that the event is instantaneous. The vast majority of GRB models (see e.g. Nemiroff 1994) are explosive: gamma-ray emitting material expands at a highly relativistic speed. We will assume they expand at the speed of light and rewrite equation (3) as $\hat{r}_e > ct_b$. Typical GRBs last a few seconds. We therefore find

$$M > 10^{-12} M_\odot \left(\frac{t_b}{1\,\mathrm{s}}\right) \left(\frac{D_{OS}}{\mathrm{Gpc}}\right)^{-1}. \qquad (7)$$

Thus a single pair of GRB spacecraft separated by 1 AU could detect or rule out a cosmological abundance of lenses over five decades of mass. Most of this range has proven inaccessible by other search techniques. A larger baseline could reach larger masses.

## 3. MASS MEASUREMENTS AND CONSTRAINTS

If GRB lensing events are detected, it will in general require additional experiments in order to estimate the mass of the lenses. However, even without additional experiments it will be possible to obtain some constraints. Understanding these constraints also allows one to extend somewhat the range of sensitivity of the experiments relative to the limits set in the previous section.

At the upper-mass limit the constraint arises from the distribution of flux ratios. Consider the opposite limit from the one examined in § 2: $a \ll \tilde{r}_e$. For this case, if one observer sees a lensed GRB, so will the other. In general the magnification difference will only be significant if the observers are fairly close to the center of the Einstein ring, $x < a/\tilde{r}_e$. In this limit, the flux ratio is given by $\mathcal{R} = (r/\tilde{r}_e)x^{-1}\cos\phi$ where $\phi$ is the angle between the spacecraft separation vector and direction to the source-lens axis. It is straight forward then to show that the distribution is given by

$$f(\mathcal{R}) = \frac{1}{2}\tau\left(\frac{a}{\tilde{r}_e}\right)^2 \mathcal{R}^{-3} \qquad F(\mathcal{R}) = \frac{1}{4}\tau\left(\frac{a}{\tilde{r}_e}\right)^2 \mathcal{R}^{-2}, \qquad (8)$$

and the same distribution for $\mathcal{R}^\dagger$. For large ratios, this distribution is almost identical to equation (4). Thus, if the experiment were sensitive only to high-ratio events, one could not distinguish between an optical depth $\tau = \tau_0$ for lenses $M < 10^{-7} M_\odot$ and a larger optical depth $\tau = 4(\tilde{r}_e/r)^2 \tau_0$ for much larger masses. However, if for example the experiment were sensitive to ratios as small as $\mathcal{R}_{\min} = 1.2$, then for the low-mass lenses, half the events would have ratios $\mathcal{R} > 1.42$, whereas for the high-mass lenses the fraction would be 71%. Hence, the two distribution could be distinguished with modest statistics.

The lower-mass limit was set by demanding that the event be shorter than the Einstein-ring crossing time. However, if the GRBs are shorter than this, lensing events will still be observable provided that the time resolution is shorter than the crossing time. In this case, the ratio of magnifications (and so fluxes) will be time-dependent with a profile exactly like that of a classical microlensing event: $\mathcal{R}(t) = A[x(t)]$ where $A(x)$ is given by equation (1), $x(t) = [\omega^2(t-t_0)^2 + \beta^2]^{1/2}$, $\omega = v/r_e$, $t_0$ is the time of maximum magnification, and $\beta$ is the dimensionless impact parameter. If $\mathcal{R}$ has no significant time dependence, this would constrain the mass $M > 10^{-12} M_\odot$. If time dependence were detected, the measured time scale would serve to characterize the size of the Einstein ring and so approximately characterize the mass. The principal uncertainty in these determinations would be the assumption that the transverse speed is $\sim c$.

If MACHOs were detected by a pair of such spacecraft, then the above constraints could be used to design new experiments which could determine the mass more precisely. If the mass were constrained to be $10^{-12} M_\odot < M < 10^{-7} M_\odot$, then a set of say five spacecraft could be launched with separations of, say, 0.3, 0.1, 0.003, 0.001, and 0.0003 AU. If the mass lay near the lower limit, then the nearest spacecraft would see similar magnifications while the most distant would see no magnification at all. If the mass were somewhat higher, several spacecraft would see similar magnifications, while the most distant would see no magnification. In this way the Einstein rings could be estimated to a factor $\sim 3$ corresponding to mass estimates of a factor $\sim 10$. If the masses were determined to be $>10^{-7}$, spacecraft could be launched with separations of, say, 3 or 10 AU. If $\mathcal{R}$ showed time dependence, then the characteristic time would already give an indication of mass. However, as noted above, the principal uncertainty would be the assumption of rapid transverse motion. It is possible, for example that the process underlying the GRB event is not relativistic, in which case $v \sim 500\,\mathrm{km\,s^{-1}}$ typical of galaxies. To distinguish between these cases, one could launch spacecraft with small separations of order the estimated size of the Einstein ring. One could then use these to measure the projected velocity, $\tilde{v} = v(\tilde{r}_e/r_e)$ in the standard manner (Gould 1994b, 1995a). The mass would then be much better constrained as would the models of GRBs.

We acknowledge comments and discussions with Will Sutherland and Tom Cline. This research was supported by grants from NASA (RJN) and the National Science Foundation AST 94-20746 (AG).